\documentclass[preprint,12pt,authoryear]{elsarticle}
\usepackage{amssymb}
\usepackage{amsmath,amscd}
\usepackage{latexsym}
\usepackage{amsthm}
\usepackage{psfrag}
\usepackage[usenames]{color}
\usepackage{textcomp}
\usepackage[all]{xy}

\newtheorem{thm}{Theorem}[section]

\newtheorem{prop}[thm]{Proposition}
\theoremstyle{definition}

\theoremstyle{remark}
\newtheorem{rem}[thm]{Remark}

\usepackage{amssymb}

\journal{Applied Mathematics Letters}

\begin{document}

\begin{frontmatter}



\title{No embedding of the  automorphisms of a topological
space into a compact metric space endows them with a composition that passes to the limit}


\author{Patrizio Frosini}

\address{Dipartimento di Matematica, Universit\`a di Bologna
}

\author{Claudia Landi}

\address{Dipartimento di Scienze e Metodi dell'Ingegneria, Universit\`a di Modena e Reggio Emilia
}

\begin{abstract}
The Hausdorff distance, the Gromov-Hausdorff, the Fr\'echet  and the natural pseudo-distances are instances of dissimilarity measures widely used in shape comparison. We show that they  share the property of being defined as $\inf_\rho F(\rho)$ where $F$ is a suitable functional and $\rho$ varies in  a set of correspondences containing the set of homeomorphisms. Our main result states that the set of homeomorphisms cannot be enlarged to a metric space $\mathcal{K}$, in such a way that the composition in $\mathcal{K}$ (extending the composition of homeomorphisms) passes to the limit and, at the same time, $\mathcal{K}$ is compact. 
\end{abstract}


\end{frontmatter}


\section{Introduction}
\label{Intro}
The literature about shape comparison often reports distances or pseudo-distances whose definitions are based on considering  sets of  correspondences between topological spaces $X$ and $Y$, where a correspondence  is defined as a surjective relation $\rho\subseteq X\times Y$ such that also $\rho^{-1}$ is surjective \citep{Me07}. In plain words, each correspondence describes a (perceptive) matching between the points of $X$ and the points of $Y$.

As a classical example, the \emph{Hausdorff distance} $d_H(X,Y)$ between two non-empty compact sets $X$ and $Y$ of a metric space $(\mathcal{S},d_{\mathcal{S}})$ is defined as the value $\inf_{\rho\in \mathcal{C}}\sup_{(x,y)\in \rho} d_{\mathcal{S}}(x,y)$, where $\mathcal{C}$ denotes the set of all correspondences between $X$ and $Y$ \citep{Me07}. The \emph{Gromov-Hausdorff pseudo-distance} \citep{BuBuIv01,Gr81} and the \emph{Fr\'echet pseudo-distance} \citep{Ro07} represent two other well-known examples where a similar procedure is applied. Sometimes (as in the case of the Fr\'echet pseudo-distance) just a proper subset of the set of all correspondences is considered.

All these examples share the property of being defined as $\inf_\rho F(\rho)$, where $F$ is  a suitable functional  taking each correspondence $\rho$ to a value that measures how much ``$\rho$ behaves as an identity'' from the point of view of our shape comparison.  In the case of the Hausdorff distance, $F(\rho)$ equals the value $\sup_{(x,y)\in \rho} d_{\mathcal{S}}(x,y)$, which vanishes if and only if $X=Y$ and $\rho$ is the identity correspondence.

In Persistent Topology the same procedure leads to the concept of \emph{natural pseudo-distance}, considering only correspondences that are also homeomorphisms. When two  closed $C^0$ manifolds $X,Y$ endowed with two continuous functions $\varphi:X\to \mathbb{R}$, $\psi:Y\to \mathbb{R}$ are considered together with the set $Hom(X,Y)$ of all homeomorphisms between $X$ and $Y$, this extended pseudo-distance is defined to be either the value $\inf_{h\in {Hom(X,Y)}}\max_{x\in X} |\varphi(x)-\psi(h(x))|$, or $+\infty$, depending on whether $X$ and $Y$ are  homeomorphic or not  \citep{FrMu99,DoFr04,DoFr07,DoFr09}.

We observe that the sets of correspondences considered in our examples    include all  homeomorphisms, which are always assumed to be legitimate transformations.


Unfortunately, in all the previous examples at least one of the following problems occurs: (1) the composition of relations does not pass to the limit; (2) the infimum of the functional $F$ is not a minimum. 
Consequently, a natural goal would be to guarantee that our functional attains a minimum by extending the metric space of the homeomorphisms between two any topological spaces $X$ and $Y$ to a compact metric space whose elements are (possibly but not necessarily) correspondences, endowed with a composition that extends the usual composition of homeomorphisms and passes to the limit.

The purpose of this paper is proving that this goal cannot be reached even in the case $X=Y$, under pretty reasonable hypotheses. This fact suggests the existence of obstacles in treating, exclusively in terms of correspondences,  the distances defined as $\inf_\rho F(\rho)$. 

\section{General setting}

Let us denote by $\mathbf{C}$ any small category (i.e. any category $\mathbf{C}$ such that both $Obj(\mathbf{C})$ and $Mor(\mathbf{C})$ are actually sets) having the following properties:

\begin{enumerate}
\item its objects are topological spaces;
\item each (possibly empty) set of morphisms $Mor(X,Y)$  between two objects $X$ and $Y$ is a subset of the set of correspondences from $X$ onto $Y$, containing all the possible homeomorphisms from $X$ onto $Y$;
\item if $\rho\in Mor(X,Y)$ then $\rho^{-1}\in Mor(Y,X)$.
\end{enumerate}

Varying $(X,Y)$ in the set $Obj(\mathbf{C})\times Obj(\mathbf{C})$, let us consider a family of  functionals $F_{(X,Y)}:Mor(X,Y)\rightarrow \mathbb{R}$ satisfying the following properties:
\begin{enumerate}
\item for every $\rho\in Mor(X,Y)$, $F_{(X,Y)}(\rho)\ge 0$;
\item if $id_X$ is the identity morphism on $X$, then $F_{(X,X)}(id_X)=0$;
\item for every $\rho\in Mor(X,Y)$, $F_{(X,Y)}(\rho)=F_{(Y,X)}(\rho^{-1})$;
\item if $\rho\in Mor(X,Y)$ and $\sigma\in Mor(Y,Z)$, $F_{(X,Z)}(\sigma\circ \rho)\le F_{(X,Y)}(\rho)+F_{(Y,Z)}(\sigma)$.
\end{enumerate}


The family of  functionals $F_{(X,Y)}$ allows us to define an extended pseudo-distance on $Obj(\mathbf{C})$ (we omit the trivial proof).  The term \emph{extended} means that the pseudo-distance can take the value $+\infty$. Obviously, passing to the quotient, any pseudo-distance becomes a distance (i.e. also the axiom $d(X,Y)=0\implies X=Y$ is satisfied).

\begin{prop}
The function
$$\delta(X,Y)=\left\{
\begin{array}{cc}
  \inf_{\rho\in Mor(X,Y)}F_{(X,Y)}(\rho) & \text{if\ } Mor(X,Y)\neq \emptyset, \\
  +\infty & \text{if\ } Mor(X,Y)= \emptyset
\end{array}
\right.$$
is an extended pseudo-distance on $Obj(\mathbf{C})$.
\end{prop}


The previous setting allows us to obtain the pseudo-distances we have recalled at the beginning of the introduction, as particular cases.\\

\noindent\textbf{Hausdorff distance}. $\mathbf{C}$ is the category whose objects are the non-empty compact subsets of a metric space $(\mathcal{S},d_{\mathcal{S}})$. The morphisms are  all  correspondences between any two objects. We set $F_{(X,Y)}(\rho)= \sup_{(x,y)\in \rho}d_{\mathcal{S}}(x,y)$, for every pair $(X,Y)\in Obj(\mathbf{C})\times Obj(\mathbf{C})$ and every $\rho\in Mor(X,Y)$.\\
  
  \noindent \textbf{Gromov-Hausdorff pseudo-distance}. $\mathbf{C}$ is a category whose objects belong to a set of non-empty compact metric spaces. The morphisms are given by all  correspondences between objects.  For every  $(X,Y)\in Obj(\mathbf{C})\times Obj(\mathbf{C})$ and  $\rho\in Mor(X,Y)$, we set $F_{(X,Y)}(\rho)= \inf_{(\mathcal{Z},d_{\mathcal{Z}}),f,g}\sup_{(x,y)\in \rho}d_{\mathcal{Z}}(f(x),g(y))$, where $(\mathcal{Z},d_{\mathcal{Z}})$ ranges over all   metric spaces, and $f$ and $g$ range over  all possible isometric embeddings of $X$ and $Y$ into $\mathcal{Z}$, respectively.\\
 
 \noindent \textbf{Fr\'echet pseudo-distance}. $\mathbf{C}$ is the category whose objects are all the curves $\gamma:[0,1]\rightarrow \mathbb{R}^n$ (seen as subsets of $[0,1]\times \mathbb{R}^n$ endowed with the product topology). The morphisms between two curves $\gamma_1,\gamma_2$ are given by the relations $\rho$ whose elements can be written as $\left(\gamma_1(\alpha(t)),\gamma_2(\beta(t))\right)$, where $t\in[0,1]$ and $\alpha,\beta:[0,1]\to [0,1]$ are two non-decreasing and surjective continuous functions. Finally, we set $F_{(X,Y)}(\rho)=\sup_{(x,y)\in \rho}\|x-y\|$.\\
 
 \noindent \textbf{Natural pseudo-distance}. $\mathbf{C}$ is the category whose objects are all the continuous functions $\varphi:X\rightarrow \mathbb{R}$, where $X$ ranges over all closed $C^0$ $n$-manifolds. They are seen as subsets of $X\times \mathbb{R}$, endowed with the product topology. The morphisms between two functions $\varphi:X\to \mathbb{R}$, $\psi:Y\to \mathbb{R}$ are given by the homeomorphisms $h$ from $X$ onto $Y$. Finally, we set $F_{(X,Y)}(h)=\max_{x\in X}|\varphi(x)-\psi(h(x))|$.

\section{Main result}

The core of this paper is the following result stating  that we cannot enlarge the set of homeomorphisms to a larger metric space $\mathcal{K}$, in such a way that the composition in $\mathcal{K}$ (extending the composition of homeomorphisms) passes to the limit and, at the same time, $\mathcal{K}$ is compact. Since the passage to the limit of the composition is important in applications because of the need for computational approximations, our result suggests that there is no sensible way to extend the set of homeomorphisms to a larger compact metric space.

\begin{thm}
\label{Result}
Let $X$ be a topological space containing a subset $U$ that is homeomorphic to an $n$-dimensional open ball for some $n\ge 1$.
Let us consider the set $\mathcal{H}$ of all homeomorphisms from $X$ onto $X$, endowed with a metric $d_\mathcal{H}$ that is compatible with the topology of $X$ in the sense of the following property: if a sequence $(h_i)$ in $\mathcal{H}$ converges to the identical homeomorphism $id_X\in \mathcal{H}$ with respect to $d_\mathcal{H}$, then $(h_i)$ pointwise converges to $id_X$ with respect to the topology of $X$ (i.e., $\lim_{i\to \infty}h_i(x)=x$ for every $x\in X$). Then no compact metric space $(\mathcal{K},d_\mathcal{K})$ exists, endowed with an internal composition
$\bullet:\mathcal{K}\times \mathcal{K}\to \mathcal{K}$ such that:
\begin{enumerate}
  \item \label{1} $\mathcal{K}\supseteq \mathcal{H}$;
  \item \label{2} $d_\mathcal{K}$ extends $d_\mathcal{H}$ (i.e. if $f,g\in \mathcal{H}$ then $d_\mathcal{K}(f,g)=d_\mathcal{H}(f,g)$);
  \item \label{3} the binary operation $\bullet$ extends the usual composition of homeomorphisms (i.e., if $f,g\in \mathcal{H}$ then $f\bullet g=f\circ g$);
  \item \label{4} the composition $\bullet$ commutes with the passage to the limit (i.e. if the sequences $(\rho_i)$ and $(\sigma_i)$ converge in $\mathcal{K}$, then $\lim_{i\to \infty}\left(\rho_i\bullet\sigma_i\right)$ exists and equals $\left(\lim_{i\to \infty}\rho_i\right)\bullet \left(\lim_{i\to \infty}\sigma_i\right)$).
\end{enumerate}
\end{thm}

\proof Let us prove our result by contradiction, assuming that such a metric space $(\mathcal{K},d_\mathcal{K})$ exists.
For every homeomorphism $f\in \mathcal{H}$ and any natural number $i> 1$, let $f^i$ denote the composition of $f$ with itself $i$ times (while $f^1=f$), and let us set $g=f^{-1}$. Since $\mathcal{K}$ is compact, a strictly increasing sequence of positive numbers $(i_r)$ exists such that both the limits, with respect to $d_\mathcal{K}$, $\lim_{r\to \infty}f^{i_r}$ and $\lim_{r\to \infty}g^{i_r}$ exist.

On one hand, if in the metric space $(\mathcal{K},d_\mathcal{K})$ we consider the constant sequence $\left(f^{i_r}\circ g^{i_r}\right)=(id_X)$, from Properties 3 and 4 it follows that
\begin{eqnarray*}
id_X&=&\lim_{r\to \infty}\left(f^{i_r}\circ g^{i_r}\right)=\lim_{r\to \infty}\left(f^{i_r}\bullet g^{i_r}\right)=\left(\lim_{r\to \infty}f^{i_r}\right)\bullet \left(\lim_{r\to \infty}g^{i_r}\right)\\
&=&\left(\lim_{r\to \infty}f^{i_{r+1}}\right)\bullet \left(\lim_{r\to \infty}g^{i_r}\right)
  =\lim_{r\to \infty}\left(f^{i_{r+1}}\bullet g^{i_r}\right)=\lim_{r\to \infty}\left(f^{i_{r+1}}\circ g^{i_r}\right).
\end{eqnarray*}

Therefore, recalling Properties 1 and 2, we have that the sequence of homeomorphisms $\left(f^{i_{r+1}}\circ g^{i_r}\right)=\left(f^{i_{r+1}-i_r}\right)$ converges to the identical homeomorphism, with respect to both $d_\mathcal{H}$ and $d_\mathcal{K}$. We observe that each index $i_{r+1}-i_r$ is strictly positive.

In other words, we have proved that for every homeomorphism $f$ from $X$ onto $X$ a sequence of positive numbers $(m_r)$ exists, such that $(f^{m_r})$ converges to the identical homeomorphism with respect to $d_\mathcal{H}$.

In order to obtain a contradiction, it is sufficient to construct a homeomorphism $h$ that cannot verify the previous property. We can do that by considering a homeomorphism $\tilde h:U\to B_n=\{x\in \mathbb{R}^n:\|x\|\le 1\}$, and constructing a homeomorphism $h\in \mathcal{H}$ that takes the set $\tilde h^{-1}\left(\{x\in \mathbb{R}^n:\|x\|\le \frac{1}{2}\}\right)$ into the set $\tilde h^{-1}\left(\{x\in \mathbb{R}^n:\|x\|\le \frac{1}{4}\}\right)$. It is immediate to check that no sequence of non-trivial positive powers of $h$ can pointwise converge to the identical homeomorphism. Therefore, no such a sequence can converge to the identical homeomorphism with respect to $d_\mathcal{H}$. \qed
\vskip 0.5cm

\begin{rem}
\label{r1}
The assumption that $X$ contains a subset $U$ that is homeomorphic to an $n$-dimensional open ball for some $n\ge 1$ cannot be omitted. Indeed, topological spaces for which the only automorphism of $X$ is the identity map exist. In that case $(\mathcal{H},d_\mathcal{H})$ is obviously compact. A classical reference for these spaces (called \emph{rigid topological spaces}) is \cite{GrWi}.
\end{rem}


\begin{rem}
\label{r3}
An important class of topological spaces for which our theorem holds is given by the triangulable spaces (i.e. the bodies of simplicial complexes) of dimension larger than or equal to $1$.
\end{rem}





\medskip

\textbf{Acknowledgement.} We thank Francesca Cagliari for her valuable suggestions.

\end{document}